\documentclass[%
 aip,
 jmp,%
 amsmath,amssymb,
 reprint,%
 unsortedaddress
]{revtex4-1}

\usepackage{graphicx}
\usepackage{bm}
\usepackage{multirow}
\usepackage{lmodern} 

\newcommand{\normShort}[1]{\lVert#1\rVert}
\newcommand{\norm}[1]{\left\lVert#1\right\rVert}
\newcommand{\pderiv}[2]{\frac{\partial #1}{\partial #2}}
\newcommand{\inlinepderiv}[2]{\partial #1 / \partial #2}

\newcommand{\Uint}{\int_{u_\mathrm{start}}^{u_\mathrm{end}}}
\newcommand*\dif{\mathop{}\!\mathrm{d}}

\newcommand{\Uvec}{\boldsymbol{U}}

\newcommand{\Pvec}{\boldsymbol{P}}

\newcommand{\Nbas}{\mathcal{N}}
\newcommand{\Pik}{P_{i,k}}
\newcommand{\Curve}{\boldsymbol{\mathcal{C}}}

\newcommand{\Hartree}{E_\mathrm{h}}
\newcommand{\Bohr}{r_\mathrm{Bohr}}
\newcommand{\eRef}{$E_\mathrm{ref}$}
\newcommand{\nRDmath}{n_\mathrm{it}}
\newcommand{\nRD}{$\nRDmath$}
\newcommand{\eRD}{$E_\mathrm{ReaDuct}$}

\newcommand{\alphac}{\alpha_\mathrm{c}}
\newcommand{\alphaf}{\alpha_\mathrm{f}}

\newcommand{\uppernote}[1]{$^{[\mathrm{#1}]}$}

\begin{document}

\preprint{AIP/123-QED}

\title{Minimum Energy Paths and Transition States by Curve Optimization}

\author{Alain C. Vaucher}
\author{Markus Reiher}
\email[Corresponding author: ]{markus.reiher@phys.chem.ethz.ch}
\affiliation{
ETH Z\"urich, Laboratorium f\"ur Physikalische Chemie, Vladimir-Prelog-Weg 2, CH-8093 Z\"urich, Switzerland
}

\date{\today}

\begin{abstract}
  Transition states and minimum energy paths are essential to understand and predict chemical reactivity.
  Double-ended methods represent a standard approach for their determination.
  We introduce a new double-ended method that optimizes reaction paths described by curves.
  Unlike other methods, our approach optimizes the curve parameters rather than distinct structures along the path.
  With molecular paths represented as continuous curves, the optimization can benefit from the advantages of an integral-based formulation.
  We call this approach ReaDuct and demonstrate its applicability for molecular paths parametrized by B-spline curves.
\end{abstract}

\maketitle

\setlength{\parindent}{0cm}
\setlength{\parskip}{0.6em plus0.2em minus0.1em}
\setlength{\tabcolsep}{3pt}

\section{Introduction}

To describe, understand, and design chemical reactivity, a multitude of different possible reactants and intermediates and a myriad of ways in which they can interact must be elucidated.
\cite{ross2008a,ludlow2008a,bell2011a,masters2011a,vinu2012a,simm2017a}
A key concept for this endeavor are elementary reactions that connect two reaction valleys.

To extract kinetic information of an elementary reaction, it is often sufficient to consider only one that leads to the smallest energy barrier.
The point with the highest energy along such a path is the transition state (TS).
It is a first-order saddle point on the potential energy surface (PES), for which the matrix of second partial derivatives of the electronic energy with respect to the nuclear coordinates (the Hessian) has exactly one negative eigenvalue.
If one follows the mode corresponding to the negative Hessian eigenvalue in both directions along the steepest descent direction, one approaches two valleys of stable structures, and the path so generated is called a minimum energy path (MEP).
All along a MEP, the path tangent is parallel to the nuclear gradient of the electronic energy.
There may exist multiple transition states connecting two molecular structures, and, accordingly, for each one we will have a different MEP.

Many algorithms to optimize transition states fall into the category of double-ended methods.
\cite{%
elber1987a,ayala1997a,%
peng1993a,%
jonsson1998a,henkelman2000a,henkelman2000b,trygubenko2004a,%
weinan2002a,weinan2005a,%
burger2006a,burger2007a,chaffeymillar2012a,%
peters2004a,goodrow2008a,%
behn2011a,sharada2012a,%
granot2008a,ghasemi2011a,%
plessow2013a,%
sheppard2008a,%
zimmerman2013b%
}
These methods aim to determine a path connecting reactant and product valleys through the transition state structure.
Some of the double-ended methods gradually construct the path starting from both ends until they connect, while others start from a reaction path guess and optimize it.
Examples belonging to the former category are the growing string method\cite{peters2004a} and the freezing string method,\cite{behn2011a,sharada2012a}
whereas the nudged elastic band (NEB) method\cite{jonsson1998a,henkelman2000a,henkelman2000b} and the string method\cite{weinan2002a,weinan2005a} are examples of the latter category.
The NEB method approximates the reaction path by a series of distinct structures (called beads or images) and optimizes them until they represent a good estimate of the minimum energy path.
To achieve this, each image undergoes a force made up of a spring force to ensure that all images are evenly spaced, and of a force perpendicular to the path that is projected from the quantum chemical forces acting on the atomic nuclei.
The string method describes the reaction path as a curve, which is updated iteratively.
At each step, structures are selected along the path, undergo a steepest descent iteration step, after which the curve is reparametrized to fit the new finite set of structures.

The reaction path delivered by double-ended methods is an approximation for the minimum energy path.
Accordingly, the structure along this reaction path with the highest electronic energy can be taken as an approximate transition state.
The exact transition state may then be optimized until the Hessian has exactly one negative eigenvalue.
For this optimization, one may rely on eigenvector following methods.\cite{cerjan1981a,simons1983a,baker1986a,jensen1995a,bergeler2015b}
In a recent work,\cite{zimmerman2013b} exact transition state search was integrated in the double-ended optimization formalism to obtain the exact transition state directly during a reaction path optimization.

The new algorithm ReaDuct presented in this work optimizes reaction paths described by (continuous) curves.
Unlike the NEB and string methods, however, it does not update iteratively structures along the path, but directly the curve itself as a whole (or, more precisely, its parametrization).
This ensures that the reaction path description is truly continuous rather than discrete, and allows for an optimization formalism relying on integrals along reaction paths.
Accordingly, the mathematical nature of the optimization is independent of the specific discretization and the optimization result can be improved at will by refining the integration grid.

ReaDuct is related to the splined saddle method,\cite{granot2008a,ghasemi2011a} which also relies on a parametrized curve and a minimization scheme for finding the transition state.
In the splined saddle method, however, the main goal of the optimization procedure is the transition state itself and the MEP is generated in a second step, while ReaDuct directly optimizes the full path between reactants and products.

We first introduce a general formalism for the optimization of curves in Section~\ref{sec:curve_formalism}.
Then, in Section~\ref{sec:readuct_formalism}, we detail the application of this formalism for the optimization of reaction paths.
We describe our implementation of ReaDuct in Section~\ref{sec:readuct_implementation} and study its application on various examples in Section~\ref{sec:readuct_examples}.

\section{Formalism: Curve optimization}
\label{sec:curve_formalism}

In this section we discuss two formalisms for optimizing curves with regard to some optimization target.
The term optimization target designs a general goal that drives the optimization of the curve.
The optimization of reaction paths will build on this formalism and will be discussed in the next section.

We consider a curve $\Curve(u) \in \mathbb{R}^n$ in $n$ dimensions defined on the domain $u \in \left[ u_\mathrm{start}, u_\mathrm{end} \right]$, $u \in \mathbb{R}$ and parametrized by $K$ parameters $p_i$, $1 \leq i \leq K$.
We consider scalar parameters, but note that depending on the specific curve parametrization, it may be possible to unite such parameters to $n$-dimensional parameter vectors.

To obtain the set of parameters $\{p_i\}$ that best fit the optimization target, we consider formalisms that are force- and cost-based.

\subsection{Force-based formalism}

\begin{figure}
\includegraphics[width=0.6\textwidth]{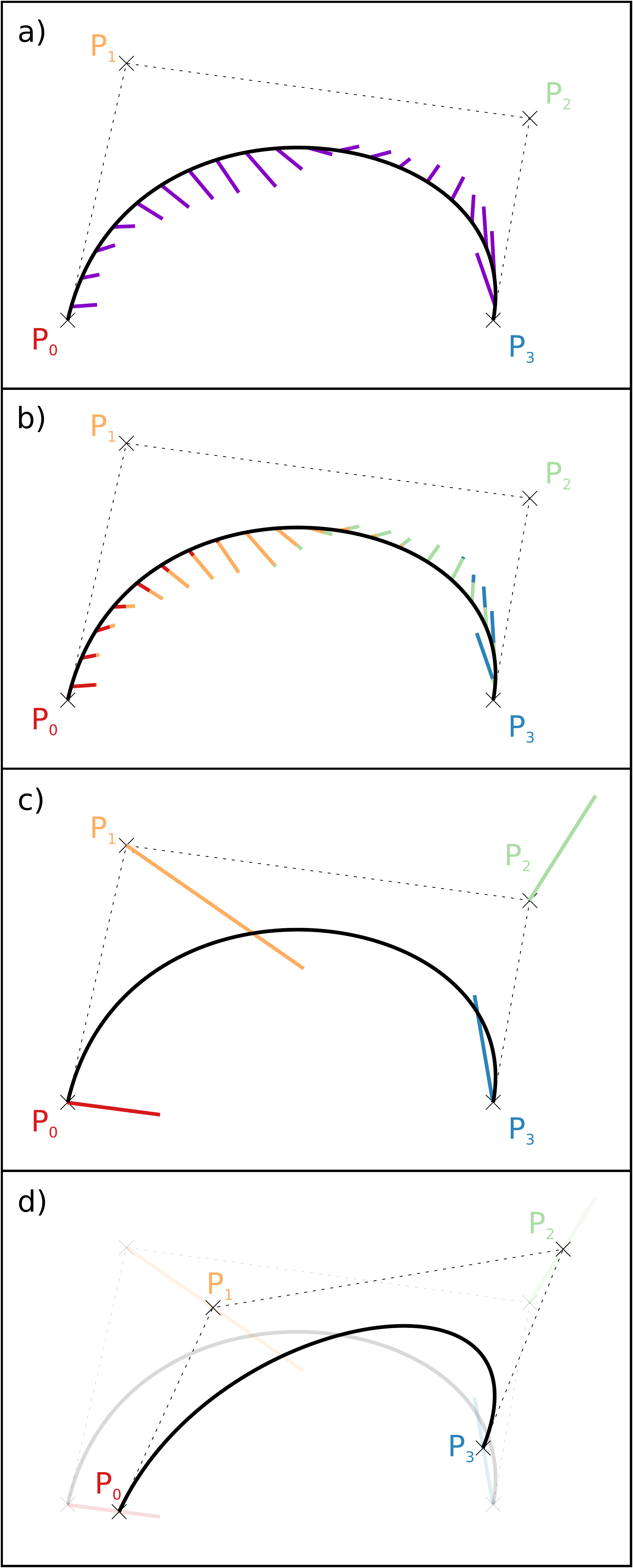}
\caption{
  Force-based optimization step of a B\'ezier curve parametrized by four two-dimensional points.
  \textbf{a)} Forces, represented by purple lines, are applied along the curve.
  \textbf{b)} Each force can be split into contributions for the different parameters, represented by colors corresponding to the parameters.
  \textbf{c)} The force for each parameter point is obtained as the sum of its contributions along the path.
  \textbf{d)} The curve is updated by changing the parameters according to the force they experience.
}
\label{fig:force_based_formalism}
\end{figure}

Given a curve, it is often straightforward to alter it in order to improve it with respect to some optimization target.
In the force-based formalism, one applies artificial forces $\boldsymbol{f}_\mathrm{local}(u) \in \mathbb{R}^n$ along the curve such that their action improves the current curve.
These artificial forces are propagated back to the curve parametrization and deliver forces $F_{\mathrm{local},i}(u)$ ($1 \leq i \leq K$) acting on the curve parameters $p_i$.
The full force on the curve parameters is then the sum of all contributions along the curve, $F_i = \Uint F_{\mathrm{local},i}(u) \dif u$.
How to propagate the forces back depends on the choice of parametrization.
The curve parameters can then be updated iteratively with an optimization algorithm such as steepest descent, FIRE,\cite{bitzek2006a} Verlet, etc.
The force-based optimization formalism is illustrated in Fig.~\ref{fig:force_based_formalism}.

\subsection{Cost-based formalism}

In a cost-based formalism, the optimization target is formulated as an objective function or cost $c$ that is to be minimized.
A cost is assigned to every possible curve, and the cost can be expressed as a function of the curve parameters.
To optimize the curve, one updates the curve parameters iteratively to decrease the cost of the curve.
When the derivatives of the cost with respect to the curve parameters are available, efficient optimization schemes can be applied for this purpose.
Examples of such schemes are the conjugate gradient method or the Broyden--Fletcher--Goldfarb--Shanno (BFGS) method.\cite{nocedal2006a}

Depending on the optimization target, one may attribute a local cost for each point along the curve, expressed as a function $c_\mathrm{local}(u)$.
The total cost is then obtained by integrating $c_\mathrm{local}$ along the curve,
\begin{align}
  c = \Uint c_\mathrm{local}(u) \dif u.
\end{align}
Then, the problem may also be formulated in the force-based formalism by choosing
\begin{align}
F_{\mathrm{local},i}(u) = - \pderiv{c_\mathrm{local}(u)}{p_i}.
\end{align}

\subsection{General Comments}

Both formalisms presented above are valid for any parametrizaton choice.
The specific parametrization will define the flexibility of the curve, and therefore the ensemble of curves that can be realized.
In general, more parameters will allow for more flexibility, and the obtained curves will fit the optimization target better.

It is important to understand that the curve is optimized as a whole by adjusting its curve parameters; one does not optimize single points along the curve.
When integrating the force or the cost along the curve, this formalism allows for improving the optimization result without changing the nature of the problem by increasing the number of integration points.

\section{Reaction path optimization}
\label{sec:readuct_formalism}

Reaction paths for molecular systems of $N$ atoms can be represented as curves in $n = 3N$ dimensions (here, we assume Cartisian coordinates; in principle, one can also work with internal coordinates).
Both the force-based and cost-based formalisms for curve optimization can be applied to reaction paths if an adequate optimization target is formulated.
The choice of a specific target may depend on the available computational resources, size of the molecular system, or desired accuracy of the optimized reaction path.
Note that the formalism presented here can also be applied on molecular paths for other purposes than the optimization of reaction paths.

We do not consider a universal choice for the optimization target the best strategy, neither in the force-based nor in the cost-based formalism.
Below, we therefore introduce two possible targets, that require only the electronic energy and its gradient,
Note, however, that the formalism also supports optimization targets depending on higher derivatives of the electronic energy.

To select an adequate optimization target, it may help to consider a two-dimensional system and regard the energy as a third dimension.
In this case, one can illustrate a reaction path by a pipe connecting two valleys separated by mountains.
In this analogy, at any time during the optimization one knows the position of the pipe, its relative height (energy), and the inclination of the landscape (gradient) along the curve.
From this information, one must decide how to alter the pipe to improve it (i.e., turn it more into a minimum energy path).

\subsection{Force-based reaction path optimization}

Given a reaction path, a reasonable direction to alter the path is downward in energy.
Therefore, in the force-based formalism a natural choice for the forces to apply along the path is the perpendicular projection of the negative gradient of the electronic energy.
For a minimal energy path, the gradient of the energy is parallel to the path and therefore no forces are applied.

To improve convergence, especially for flat parts of the potential energy surface, one can add a contribution for the path tension.

The force applied on the curve is therefore given by
\begin{align}
  \boldsymbol{f}_\mathrm{local}(u) = (1 - \alphaf) \cdot \boldsymbol{f}_{\mathrm{perp}}(u) + \alphaf \cdot \boldsymbol{f}_{\mathrm{para}}(u),
\end{align}
where $\alphaf$ is a parameter of the method (the subscript `f' relates to the force-based formalism).
The perpendicular contribution from the negative gradient of the electronic energy is given by
\begin{align}
  \boldsymbol{f}_{\mathrm{perp}}(u) = - \left( \boldsymbol{\nabla} E_\mathrm{el}(\Curve(u)) \right)_\perp \cdot \frac{\Bohr^2}{\Hartree},
\end{align}
where $\Hartree$ is the Hartree energy unit and $\Bohr$ is the Bohr radius, and the path tension contribution is parallel to the reaction path,
\begin{align}
  \boldsymbol{f}_{\mathrm{para}}(u) = 2 \cdot \left( \left( \Curve'(u) \right)^T \Curve''(u) \right) \cdot \hat{\boldsymbol{t}} \cdot \frac{1}{\Bohr},
\end{align}
where $\hat{\boldsymbol{t}}$ denotes the unit tangent.
Note that the forces defined here have the dimension of a length for practical reasons.
Since these forces have no true physical meaning, other units would be possible.
Note that this formulation is one of a multitude of possibilities.
It is closely related to the NEB method and could be considered its continous-path equivalent.

\subsection{Cost-based reaction path optimization}

To optimize a reaction path in the cost-based formalism, one must reduce its information to a single number (its cost).
This requires special care, since artefacts may arise depending on the cost formulation.

To illustrate this issue, let us consider a cost function $c_1$ equal to the integral of the energy along the reaction path.
Note that similar strategies were followed based on a discretized reaction path.\cite{elber1987a,ayala1997a}
This cost can be formulated as
\begin{align}
  c_1 = \Uint E_\mathrm{el}(\Curve(u)) \dif L,
\end{align}
where
\begin{align}
  \dif L = \norm{ \pderiv{\Curve (u)}{u} } \dif u = \norm{ \Curve'(u) } \dif u
\end{align}
is an infinitesimal path segment.
While $c_1$ may be reasonable, it is based on a multiplication of a length with an energy.
Hence, not only $c_1$, but also the position of the optimized curve depends on the energy reference.
Furthermore, minimization of $c_1$ will fail for negative energies, because in such a case the cost keeps decreasing for increasing reaction path lengths.

To remove the dependence on the energy reference, one can normalize the cost by dividing it by the total path length,
\begin{align}
  c_2 = \frac{c_1}{\displaystyle \Uint \dif L}.
\end{align}
$c_2$ will still depend on the energy reference, but the position of its minima will not.
However, another artefact is introduced: for a given reaction path, the cost can be lowered if the path length increases in low-energy regions and thereby reduces the weight of high-energy regions.

To avoid these issues for $c_1$ and $c_2$, one may define the related cost function
\begin{align}
  c_3 = \Uint E_\mathrm{el}(\Curve(u)) \dif u.
\end{align}
This formulation avoids the need for explicit normalization if the integration interval ($\left[ u_\mathrm{start}, u_\mathrm{end} \right]$) is constant.
It may, however, lead to an uneven distribution of $u$ along the curve to increase the contribution of low-energy regions.
To resolve this, one can add, similarly to the force-based reaction path optimization, a term for the curve tension:
\begin{align}
  c_\mathrm{ReaDuct} = (1 - \alphac) \cdot c_\mathrm{energy} + \alphac \cdot c_\mathrm{tension}
\end{align}
with
\begin{align}
  c_\mathrm{energy} = c_3 \cdot \frac{1}{\Hartree} = \left( \Uint E_\mathrm{el}(\Curve(u)) \dif u \right) \cdot \frac{1}{\Hartree}
\end{align}
and
\begin{align}
  c_\mathrm{tension} = \Uint \left( \pderiv{\norm{\Curve'(u)}^2}{u} \right)^2 \dif u \cdot \frac{1}{\Bohr^4},
\end{align}
where $\alphac$ is a parameter of the method (the $\mathrm{c}$ subscript relates to the cost-based formalism).
Note that we define $c_\mathrm{ReaDuct}$ to have no units.
In the following sections, all examples for the cost-based formalism will rely on $c_\mathrm{ReaDuct}$.
Note that although the gradients of the electronic energy do not appear in $c_\mathrm{ReaDuct}$, they will be required for the reaction path optimization (through the derivatives of $c_\mathrm{ReaDuct}$ with respect to the curve parameters).

\subsection{Initial reaction path}

The reaction path from which the optimization is started is crucial for the course of the reaction path optimization.
Especially, for reactions with multiple possible minimum energy paths, the choice of the initial reaction path will affect which one is obtained after optimization.
While this is not the focus of the present work, we will mention a few common choices for initial paths.

Given some structures for the reactants and products of a reaction, one can generate an initial curve by interpolating between them.
Note that the order of the atoms in the structure description must be identical in both structures (atom matching).

Another possibility to determine the initial reaction path is to fit a curve to a sequence of structures.
This allows for processing reactivity information of first-principles molecular dynamics simulations, interactive explorations of reactivity, or PES scans (see the discussion in Ref.~\citenum{heuer2018a}).

It is also possible to start from paths already available as curves, which can happen when starting from pre-processed reaction paths (see below) or from paths optimized with other quantum chemical methods.

Rotation and translation should be removed from the initial path, especially when the optimization target formulation depends on the curve tangent.
Otherwise, the cost or forces underlying the reaction path optimization may depend on the overall rotation and translation of the molecular system.
Removal of rotation and translation can for instance be achieved with the help of quaternions.\cite{coutsias2004a}
A formulation of the curve in internal coordinates avoids such issues.

\subsection{Improved initial reaction path}

The initial reaction path may be very different from the minimal energy path.
This is often the case when the initial reaction path is generated by interpolation between start and end structures.
In this case, it may be useful to improve this guess in order to reduce the number of single-point calculations required for the reaction path optimization, especially if this is possible without much computational effort (e.g., without explicit single-point calculations).

In our formalism, improvement can be achieved naturally by optimizing the path with a cost function $c_\mathrm{improved}$ that is independent of the electronic structure.
One possibility is related to the image-dependent pair potential (IDPP) for initial paths in the NEB method.\cite{smidstrup2014a}
IDPP defines, for each image along the NEB path, a target distance for all pairs of atomic nuclei and an objective function that aims to match the target distances as closely as possible.
We proceed similarly but define such an objective function along the (continuous) reaction path.
The minimization of this cost function is naturally supported by the formalism detailed in the previous section.
Hence, we define the cost for an improved initial reaction path as the integral along the path of the sum of pair potentials for the atomic nuclei,
\begin{align}
  c_\mathrm{improved} = \Uint \sum_i^N \sum_{j>i}^N  V \left( r_{ij}(u), r_{\mathrm{target},ij}(u) \right) \dif u,
\end{align}
where $r_{ij}(u)$ is the distance between atomic nuclei $i$ and $j$ at the curve coordinate $u$.
It can be extracted from $\Curve(u)$.
$r_{\mathrm{target},ij}(u)$ is the target distance for this atom pair at $u$ and can be defined on the domain $ \left[ u_\mathrm{start},u_\mathrm{end} \right] $ as a linear interpolation between the internuclear distance at the start and end points of the reaction path,
\begin{align}
  r_{\mathrm{target},ij}(u) = \frac{u \cdot r_{ij}(u_{\mathrm{end}}) + (u_{\mathrm{end}} - u) \cdot r_{ij}(u_{\mathrm{start}})}{u_{\mathrm{end}} - u_{\mathrm{start}}}.
\end{align}
Different choices can be adopted for the pair potential $V(r, r_\mathrm{target})$.
For instance, the IDPP pair potential may be formulated as
\begin{align}
  V_\mathrm{IDPP}(r, r_\mathrm{target}) = \frac{1}{r_\mathrm{target}^4} \left( r_\mathrm{target} - r \right)^2 \cdot \Bohr^2.
\end{align}

\section{Implementation}
\label{sec:readuct_implementation}

We implemented the ReaDuct algorithm as a module of our software package \texttt{SCINE}.\cite{scine}
\texttt{SCINE} provides implementations of the Parametrized Method 6 (PM6),\cite{stewart2007} the non-self-consistent density-functional tight-binding (DFTB) method,\cite{porezag1995,seifert1996} as well as DFTB2\cite{elstner1998} and DFTB3.\cite{gaus2011}
These semiempirical methods enable reaction path optimization of molecular systems of moderate size in a few seconds.
\texttt{SCINE} also offers the possibility to steer the quantum chemical programs \texttt{ORCA},\cite{neese2012a} \texttt{Gaussian}\cite{gaussian16} and \texttt{Q-Chem}\cite{shao2015a} so that more advanced electronic structure methods are also available

We represent reaction paths as cubic B-spline curves.
B-spline curves are functions $\Curve(u)$ composed of $n$ piecewise polynomial segments of degree $p$, whereby $1\leq p \leq n$.
A B-spline curve is given by
\begin{align}
\Curve(u) = \sum_{i=0}^{n} \Nbas_{i,p}^{\Uvec} (u)\, \Pvec_i,
\end{align}
where $\Nbas_{i,p}^{\Uvec} (u)$ are B-spline functions defined on a so-called knot vector $\Uvec$.
$\Pvec_i$ is called control point vector.
B-Spline curves are commonly defined on the interval $u \in \left[0, 1\right]$.
A detailed discussion can be found elsewhere.\cite{piegl1997a,heuer2018a}

The B-spline parameters subject to optimization are the coordinates of the control points.
It is important to note that although the control points have the same dimension as structures along the curve, they should not be interpreted as such.
In general, we do not optimize the first and last control points, which correspond to the two end points of the path.

In Appendix~\ref{appendix:derivative_calculation}, we illustrate how the derivatives of a cost function with respect to the control point coordinates can be calculated.

The applicability of ReaDuct is not limited to specific methods, and is applicable as long as energies and gradients are available.
Apart from quantum chemical methods, molecular mechanics approaches can also be applied.

During reaction path optimization, the single-point calculations are accelerated with adequate initial electronic densities to reduce the number of self-consistent field iterations.
This is currently achieved by storing the electronic density from one step to the next for each integration point along the path, but propagation approaches\cite{atsumi2008,atsumi2010,muehlbach2016a} are also possible.
Non-linear optimization of the curve cost relies on algorithms implemented in the \texttt{dlib} library.\cite{king2009a}

\begin{figure}
\includegraphics[width=0.7\textwidth]{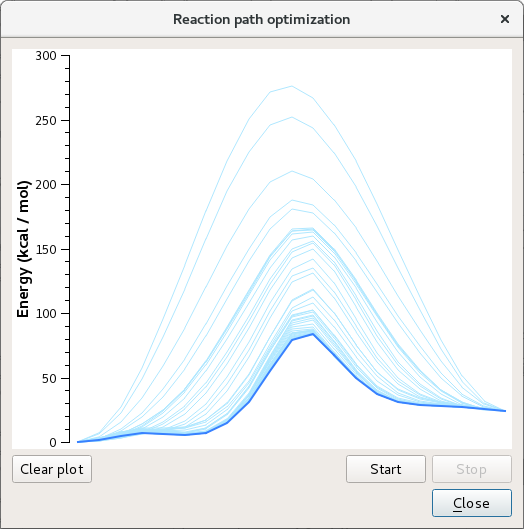}
\caption{
  Window shown during a reaction path optimization.
  Every energy profile represents a reaction path visited during the optimization.
  The energy profile represented by the thickest line represents the current (most recent) reaction path.
}
\label{fig:window_energy_profiles}
\end{figure}

\begin{figure}
\includegraphics[width=0.7\textwidth]{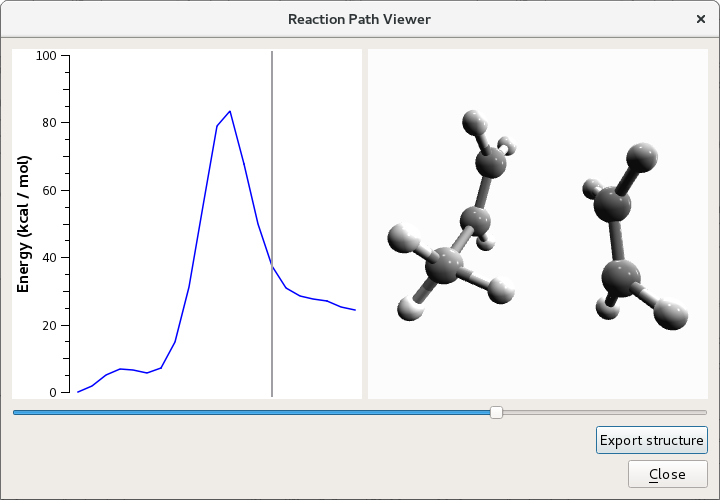}
\caption{
  Window shown to visualize a reaction profile.
  A slider bar allows one to follow the reaction coordinate and show the corresponding molecular structures.
}
\label{fig:window_view_profile}
\end{figure}

\texttt{SCINE} is available in a graphical (window-based) and in a non-graphical variant.
The graphical version allows for reaction path optimizations in real time and is especially adequate in combination with semiempirical methods.
It shows the evolution of energy profiles during the optimization (Fig.~\ref{fig:window_energy_profiles}) and the built-in molecular viewer illustrates the reaction path reaction coordinate (Fig.~\ref{fig:window_view_profile}).

The non-graphical version is designed to be executed on the command line, possibly on high-performance-computing facilities.
The non-graphical version of \texttt{SCINE} is more practical for automated reaction path optimizations such as those described in Refs.\ \citenum{bergeler2015a} and \citenum{simm2017a}, large numbers of reaction path optimizations, or for more time-demanding single-point calculations.
It reads in a YAML\cite{yaml12}-formatted input file that describes the different tasks to performed.
The input file can be generated interactively with the graphical program version.
It is also possible to visualize the reaction paths calculated non-graphically in the graphical program version.

For analysis purposes of the obtained MEPs and transition states, \texttt{SCINE} also implements an eigenvector following method\cite{baker1986a} and an intrinsic reaction coordinate following algorithm.

\section{Examples}
\label{sec:readuct_examples}

To asses the reliability of the ReaDuct algorithm, we studied an example set of 20 reactions discussed in a related work about transition state optimization by Birkholz and Schlegel.\cite{birkholz2015a}
The interested reader can find a list of the reactions in Fig.\ 1 of Ref.\ \citenum{birkholz2015a}.
Below, the reaction labels will also refer to the ones defined in that figure of Ref.\ \citenum{birkholz2015a}.
Unless stated otherwise, the calculations refer to PM6 results.

In the Supporting Information, we discuss the choice and importance of the different ReaDuct parameters at the example set.
There, we consider both force-based and cost-based formulations of ReaDuct, and show that cost-based optimizations require significantly fewer steps to obtain reasonable transition state candidates.
Therefore, we focus on cost-based optimizations here.

The examples discussed below were all obtained with the same set of parameters.
The optimized reaction paths were described by B-spline curves with five control points, and the integrations involved in the optimization considered 11 equidistant points along the reaction paths at uniformly distributed curve coordinates.
A value of $10^{-5}$ was chosen for $\alphac$, and the optimizations were stopped when the root mean square of the cost derivatives with respect to the curve parameters (control point coordinates) fell below $10^{-3} \Bohr^{-1}$.
As can be seen in the Supporting Information, this threshold is adequate for a fast determination of transition state candidates.
To improve on the transition state candidates, it is possible to choose a tighter threshold (but note that this will lead to a larger number of optimization steps and, accordingly, to an increased computational time).
The BFGS optimization algorithm was used for the cost minimization.
When applicable, the improved initial path was generated with the IDPP cost function in an optimization with 81 integration points, and a threshold for the root mean square of the cost derivatives of $10^{-5} \Bohr^{-1}$.

\subsection{Application of ReaDuct from reactant and product structures}

One can generate an initial reaction path by interpolating the coordinates given in the Supporting Information of Ref.\ \citenum{birkholz2015a}.
The results obtained with ReaDuct are shown in Table~\ref{tab:readuct_interpolation_cost} for the cost-based formalism, with (IDPP, PM6) and without (PM6) improved initial path with the IDPP cost function.

For most cases, the transition state candidates delivered by ReaDuct are good approximations to the desired transition state.
Exceptions are the Cope and OxyCope reactions, where other transition states with higher energies were found.
In both cases, a better initial reaction path, such as the one discussed in the Supporting Information, leads to the correct transition state.
The application of the IDPP cost function does not make a significant difference, except for cases in which the linear interpolation was particularly inadequate such as for the Cope, HCN, and OxyCope reactions.

\begin{table}[tp]
\centering
\caption{
  Results of the application of ReaDuct to the test reactions, starting from an interpolation between the structures for the reactants and products, in the cost-based formalism.
  \eRef{} refers to the energy of the transition state structure given in Ref.\ \citenum{birkholz2015a}, in $\Hartree$.
  \eRD{} is the energy of the transition state candidate delivered by ReaDuct in $\Hartree$.
  \nRD{} refers to the number of iteration steps in the optimization procedure, and can be multiplied by the number of integration points (here, 11) to determine the total number of single-point calculations.
}
\label{tab:readuct_interpolation_cost}
\begin{tabular}{c|c|cc|cc}
\hline\hline
          &          & \multicolumn{2}{|c|}{PM6} & \multicolumn{2}{|c}{IDPP, PM6} \\
Reaction  & \eRef    & \nRD  & \eRD      & \nRD    & \eRD        \\
\hline
C$_2$N$_2$O     & -34.7508  & 10  & -34.7492  & 9   & -34.7489  \\
C$_5$HT      & -26.4037  & 19  & -26.3938  & 28  & -26.3935  \\
Cope      & -31.9092  & 208 & -31.6953  & 86  & -31.7216  \\
CPHT      & -25.3950  & 6   & -25.3899  & 10  & -25.3880  \\
Cyc-But   & -20.8842  & 10  & -20.8772  & 29  & -20.8796  \\
DACP2     & -50.8593  & 18  & -50.8394  & 43  & -50.8419  \\
DACP+eth & -36.3806  & 8   & -36.3713  & 12  & -36.3720  \\
DFCP      & -50.0347  & 24  & -50.0136  & 32  & -50.0159  \\
Ene       & -27.3982  & 31  & -27.3522  & 60  & -27.3536  \\
Grignard  & -112.9261 & 32  & -112.9016 & 35  & -112.8954 \\
H$_2$+CO    & -16.0850  & 18  & -16.0831  & 21  & -16.0833  \\
HCN       & -11.4217  & 477 & -11.4146  & 17  & -11.4202  \\
HF+eth   & -28.7245  & 14  & -28.7170  & 27  & -28.7178  \\
Hydro     & -66.8948  & 45  & -66.8368  & 69  & -66.8226  \\
MeOH      & -17.0795  & 30  & -17.0437  & 43  & -17.0436  \\
Oxirane   & -88.1479  & 17  & -88.1376  & 31  & -88.1442  \\
OxyCope   & -37.1128  & 361 & -36.9411  & 105 & -36.9472  \\
Silane    & -4.5768   & 18  & -4.5583   & 22  & -4.5583   \\
S$_\mathrm{N}$2       & -32.7061  & 10  & -32.7032  & 13  & -32.7035  \\
Sulfolene & -48.7413  & 36  & -48.7231  & 56  & -48.7194 \\
\hline\hline
\end{tabular}
\end{table}

\subsection{Application of ReaDuct from sequences of structures}

To have more control over the specific local transition states, the initial path can be generated from a sequence of structures corresponding roughly to the MEP of interest.
We generated a set of 20 initial paths for the test reactions by conducting the reactions interactively\cite{haag2014b} in the \texttt{SCINE} extension for real-time quantum chemistry.
The corresponding sequences of structures are available in the Supporting Information.
B-spline curves were fit to the so generated sequences of structures, and a geometry optimization made sure that the initial reaction paths started and ended at stable structures.

Table~\ref{tab:readuct_haptic_cost} summarizes the optimization results.
In all cases, the desired transition state was found.
Starting the optimization from an improved guess with the IDPP cost function is beneficial in all cases and reduces the number of optimization steps.
This observation can be ascribed to the fact that IDPP-optimized paths correspond to smoother transitions from reactants to products that are closer to the MEPs.

\begin{table}[tp]
\centering
\caption{
  Results of the application of ReaDuct to the test reactions, starting from sequences of structures generated in a haptic exploration, in the cost-based formalism.
  \eRef{} refers to the energy of the transition state structure given in Ref.\ \citenum{birkholz2015a}, in $\Hartree$.
  \eRD{} is the energy of the transition state candidate delivered by ReaDuct in $\Hartree$.
  \nRD{} refers to the number of iteration steps in the optimization procedure.
}
\label{tab:readuct_haptic_cost}
\begin{tabular}{c|c|cc|cc}
\hline\hline
          &          & \multicolumn{2}{|c|}{PM6} & \multicolumn{2}{|c}{IDPP, PM6} \\
Reaction  & \eRef    & \nRD  & \eRD      & \nRD    & \eRD        \\
\hline
C$_2$N$_2$O     & -34.7508  & 39  & -34.7476  & 11 & -34.7481  \\
C$_5$HT      & -26.4037  & 61  & -26.3897  & 29 & -26.3899  \\
Cope      & -31.9092  & 119 & -31.7977  & 95 & -31.8031  \\
CPHT      & -25.3950  & 18  & -25.3908  & 13 & -25.3912  \\
Cyc-But   & -20.8842  & 46  & -20.8787  & 35 & -20.8796  \\
DACP2     & -50.8593  & 66  & -50.8473  & 24 & -50.8436  \\
DACP+eth & -36.3806  & 39  & -36.3710  & 18 & -36.3691  \\
DFCP      & -50.0347  & 13  & -50.0314  & 12 & -50.0304  \\
Ene       & -27.3982  & 91  & -27.3440  & 62 & -27.3511  \\
Grignard  & -112.9261 & 108 & -112.8926 & 27 & -112.8986 \\
H$_2$+CO    & -16.0850  & 46  & -16.0844  & 21 & -16.0603  \\
HCN       & -11.4217  & 24  & -11.4216  & 12 & -11.4212  \\
HF+eth   & -28.7245  & 51  & -28.7198  & 25 & -28.7195  \\
Hydro     & -66.8948  & 89  & -66.8498  & 61 & -66.8482  \\
MeOH      & -17.0795  & 33  & -17.0719  & 23 & -17.0668  \\
Oxirane   & -88.1479  & 89  & -88.1421  & 32 & -88.1454  \\
OxyCope   & -37.1128  & 91  & -37.0599  & 71 & -37.0682  \\
Silane    & -4.5768   & 19  & -4.5759   & 11 & -4.5761   \\
S$_\mathrm{N}$2       & -32.7061  & 20  & -32.7030  & 10 & -32.7042  \\
Sulfolene & -48.7413  & 51  & -48.7013  & 71 & -48.7052 \\
\hline\hline
\end{tabular}
\end{table}

\subsection{Density Functional Theory}

The application of ReaDuct to the example set was studied for density funtional theory (DFT).
For the DFT single-point calculations, \texttt{SCINE} steered the \texttt{Gaussian} program package.\cite{gaussian16}
The coordinates for reactant, product, and reference transition state structures were taken from Ref.~\citenum{birkholz2015a}.
Table~\ref{tab:readuct_gaussian} summarizes the B3LYP/6-31G(d,p) results.
It is also possible to start DFT reaction path optimizations with initial paths obtained from path optimizations with PM6.
We studied this for the example set, and the results are also given in Table~\ref{tab:readuct_gaussian}.

Some of the optimizations failed due to inadequate initial reaction paths or nonconverging calculations.
These cases are flagged in Table~\ref{tab:readuct_gaussian}.
In all cases, starting from the PM6-optimized path reduced considerably the number of optimization steps.

\begin{table*}[tp]
\centering
\caption{
  Results of the application of ReaDuct to the test reactions with B3LYP/6-31G(d,p).
  \eRef{} refers to the energy of the transition state structure given in Ref.\ \citenum{birkholz2015a}, in $\Hartree$.
  \eRD{} is the energy of the transition state candidate delivered by ReaDuct in $\Hartree$.
  \nRD{} refers to the number of iteration steps in the optimization procedure.
  Comments:
  [a] Unsuccesful because of nuclear overlap in the linear interpolation.
  [b] Unsuccesful because of inadequate IDPP guess.
  [c] SCF Calculations not converging during optimization.
  [d] Results obtained with the \texttt{Gaussian} flag "SCF=XQC".
}
\label{tab:readuct_gaussian}
\begin{tabular}{c|c|cc|cc|cc|cc}
\hline\hline
          &          & \multicolumn{2}{|c|}{DFT} & \multicolumn{2}{|c|}{IDPP, DFT} & \multicolumn{2}{|c|}{PM6, DFT} & \multicolumn{2}{|c}{IDPP, PM6, DFT} \\
Reaction  & \eRef    & \nRD  & \eRD                     & \nRD  & \eRD                     & \nRD  & \eRD                     & \nRD    & \eRD        \\
\hline
C$_2$N$_2$O     & -263.2163  & 12     & -263.2153 & 9      & -263.2153 & 3      & -263.2149 & 3      & -263.2150 \\
C$_5$HT      & -195.2614  & 15     & -195.2561 & 21     & -195.2560 & 3      & -195.2542 & 5      & -195.2546 \\
Cope      & -234.5720  & 44     & -234.5118 & 73     & -234.5129 & 14     & -234.5070 & 17     & -234.5087 \\
CPHT      & -194.0664  & 10     & -194.0611 & 12     & -194.0590 & 3      & -194.0547 & 3      & -194.0547 \\
Cyc-But   & -155.9262  & 12     & -155.9181 & 41     & -155.9246 & 6      & -155.9188 & 5      & -155.9198 \\
DACP2     & -388.1902  & 14     & -388.1681 & 35     & -388.1746 & 8      & -388.1727 & 3      & -388.1739 \\
DACP+eth & -272.6725  & 13     & -272.6642 & 17     & -272.6652 & 4      & -272.6632 & 4      & -272.6662 \\
DFCP      & -316.2708  & 9      & -316.2609 & 10     & -316.2593 & 6      & -316.2601 & 4      & -316.2568 \\
Ene       & -196.4582  & 32     & -196.4215 & 54     & -196.3996 & 10     & -196.4200 & 8      & -196.4217 \\
Grignard  & -3580.1803 & 22    & -3580.1423 & 40     & -3580.1430 & 17    & -3580.1437 & 9     & -3580.1393 \\
H$_2$+CO    & -114.3649  & 24     & -114.3623 & 18     & -114.2560 & 8      & -114.3597 & 9      & -114.3560 \\
HCN       & -93.3474   & Fail\uppernote{a} & Fail\uppernote{a}    & Fail\uppernote{b} & Fail\uppernote{b}    & 10     & -93.3360  & 9      & -93.3356  \\
HF+eth   & -178.9610  & 19     & -178.9582 & 32     & -178.9588 & 10     & -178.9577 & 9      & -178.9569 \\
Hydro     & -460.5243  & 39     & -460.4599 & 92     & -460.4647 & 24     & -460.4661 & 19     & -460.4610 \\
MeOH      & -115.5732  & 21     & -115.5442 & 42     & -115.5457 & 9      & -115.5316 & 10     & -115.5327 \\
Oxirane   & -652.6848  & 26     & -652.6675 & 40     & -652.6712 & 8      & -652.6651 & 9      & -652.6711 \\
OxyCope   & -270.4669  & Fail\uppernote{a} & Fail\uppernote{a}    & Fail\uppernote{c} & Fail\uppernote{c}  & 22     & -270.2363 & 25     & -270.0815 \\
Silane    & -291.7942  & 32     & -291.7674 & 51     & -291.7656 & 25\uppernote{d}     & -291.7638\uppernote{d} & 28\uppernote{d}     & -291.7640\uppernote{d} \\
S$_\mathrm{N}$2       & -717.9007  & 14     & -717.8972 & 27     & -717.8978 & 11     & -717.8991 & 6      & -717.8992 \\
Sulfolene & -704.5600  & 16     & -704.5416 & 36     & -704.5465 & 11     & -704.5495 & 11     & -704.5468 \\
\hline\hline
\end{tabular}
\end{table*}

\subsection{Video examples}

Videos illustrating the real-time optimization of reaction paths can be found on the Internet.\cite{reaductVideos}
They illustrate the following:
\begin{itemize}
  \item Cost-based optimization of the DACP+eth reaction (Diels--Alder reaction of cyclopentadiene and ethylene), followed by eigenvector following and normal mode analysis.
  \item Effect of the number of integration points for the optimization of the Grignard reaction (addition of phenyl magnesium bromide to benzophenone).
  \item Occurrence and cure of incorrect SCF convergence\cite{vaucher2017a} during reaction path optimizations at the example of the Silane reaction (addition of silylene to H$_2$).
  \item How to find different transition states for the Cope reaction (Cope rearrangement of 1,4-hexadiene) by optimizing paths recorded in interactive explorations of chemical reactivity.
\end{itemize}

\section{Conclusion and Outlook}

In this work, we introduced the ReaDuct algorithm for the optimization of reaction paths and the search of transition states.
ReaDuct is a double-ended method that relies on the optimization of curves rather than single structures along a path.
This is achieved by updating the curve parameters directly.

We presented two formalisms to carry out an iterative optimization of a reaction path.
Both aim at finding the curve that delivers the best fit to some given optimization target.
In the force-based formalism, one defines artificial forces along the curve at every iteration step.
These forces are propagated back to the curve parameters, which are then updated to alter the curve.
In the cost-based formalism, the goal is to optimize the curve with respect to an objective function, or cost, expressed as a function of the curve parameters.
The cost can be minimized by application of standard minimization algorithms.

For both formalisms, in the most common scenario the objective function can be defined locally for each point along the curve.
Then, the optimization problem involves an integral of some function along the curve.
This offers a clean formulation that is physically consistent for infinitesimally small integration steps, which is not the case for related (discrete) double-ended string methods.
Consequently, this clean formalism features the advantage to be arbitrarily refineable and extensible.

The initial reaction path extensively affects the number of iteration steps and the transition state found.
Possible initial reaction paths are interpolations between reactant and product structures or sequences of structures resulting from first-principle molecular dynamics simulations, reaction scans, or interactive reactivity explorations.
The ReaDuct formalism naturally supports the generation of improved initial reaction paths with objective functions that do not require additional single-point calculations.

We implemented the ReaDuct formalism in our cross-platform quantum chemistry package \texttt{SCINE}.
In our current implementation, reaction paths are represented by cubic B-spline curves.
Reaction paths can be optimized with semiempirical methods available in \texttt{SCINE} or with other quantum chemical methods relying on interfaces to different quantum chemistry packages.
In addition to a terminal-based version, \texttt{SCINE} offers a graphical user interface with built-in molecular viewer that allows for optimization of reaction paths in real time.

While the results achieved so far already illustrate the merits of the ReaDuct formalism, the computational cost of the method and quality of the optimized reaction paths can still be improved in different ways.
First, there may be other optimization targets than the two we introduced in this work, which better fit the optimization of reaction paths.
Also, new targets for improved initial paths may further reduce the number of optimization steps and improve the success rate of ReaDuct.
Second, our present implementation integrates the curve with an equidistant grid at every optimization step.
It may be computationally beneficial to do this following an importance sampling or to slowly freeze the end points of the curve after the first iterations analogously to what is done in the freezing string method.
Third, witnessing the importance of the initial reaction path, it may be worth designing new approaches for their determination to ensure that the desired transition states are found and to further reduce the number of optimization steps.
Fourth, integrated approaches\cite{zimmerman2013b} for the determination of the exact transition state may allow for a faster determination of transition states.

These directions are only four possibilities among others.
The ReaDuct formalism is general and extensible enough for a wide range of application in computational chemistry, and this work represents the first step in this direction.

\section*{Acknowledgments}

We gratefully acknowledge generous support by ETH Zurich (grant number: ETH-20 15-1).

\section*{Appendix}
\setcounter{section}{0}
\renewcommand{\thesection}{\Alph{section}}

\section{Cost derivative calculation}\label{appendix:derivative_calculation}

In the following, we detail the calculation of the cost derivatives with respect to the parametrization of a B-spline curve.
In this work, the knot vector $\Uvec$ of B-spline curves was not optimized, 
only the control points, $\Pvec_i$, were.
The calculation of the derivative of the cost $c$ with respect to the $k$-th coordinate of the $i$-th control point, $\Pik$, is shown
for one choice of a cost function.

Let us consider the cost function as an integral of the electronic energy along the reaction path,
\begin{align}
c = \Uint E_\mathrm{el}(\Curve(u)) \dif L.
\end{align}

We identify
\begin{align}
  \dif L = \norm{ \pderiv{\Curve (u)}{u} } \dif u = \norm{ \Curve'(u) } \dif u
\end{align}
to write
\begin{align}
c = \Uint E_\mathrm{el}(\Curve(u)) \norm{ \Curve'(u) } \dif u
\end{align}

The derivative with respect to any curve parameter $\Pik$ ($k$-th component of the control point $\boldsymbol{P}_i$) is then given by
\begin{align}
\pderiv{c}{\Pik} &= \pderiv{}{\Pik} \left( \Uint E_\mathrm{el}(\Curve(u)) \norm{ \Curve'(u) } \dif u \right) \nonumber \\
                  &= \Uint \left( \pderiv{}{\Pik} \left( E_\mathrm{el}(\Curve(u)) \normShort{ \Curve'(u) } \right) \right) \dif u \nonumber \\
                  &= \Uint \left( \pderiv{E_\mathrm{el}(\Curve(u))}{\Pik} \normShort{ \Curve'(u) } + E_\mathrm{el}(\Curve(u)) \pderiv{\normShort{ \Curve'(u) }}{\Pik} \right) \dif u
\label{eq:full_derivative_example}
\end{align}

In the first term of Eq.\ (\ref{eq:full_derivative_example}), we identify
\begin{align}
  \pderiv{E_\mathrm{el}(\Curve(u))}{\Pik} &= \sum_{j=1}^{3N} \pderiv{E_\mathrm{el}(\Curve(u))}{C_j(u)} \pderiv{C_j(u)}{\Pik} \nonumber \\
                       &= \pderiv{E_\mathrm{el}(\Curve(u))}{C_k(u)} \pderiv{C_k(u)}{\Pik}
\end{align}
where the sum collapses because $\Pik$ only affects the $k$-th dimension, $C_k$, of the B-spline curve.
$\inlinepderiv{E_\mathrm{el}(\Curve(u))}{C_k(u)}$ corresponds to the $k$-th dimension of the energy gradient at the coordinate $u$ along the spline.
$\inlinepderiv{C_k(u)}{\Pik}$ can be calculated with standard algorithms.\cite{heuer2018a,piegl1997a}

The second term of Eq.\ (\ref{eq:full_derivative_example}) is obtained as
\begin{align}
  \pderiv{}{\Pik} \normShort{ \Curve'(u) }&= \pderiv{}{\Pik} \sqrt{\normShort{\Curve'(u)}^2} \nonumber \\
                                           &= \pderiv{}{\Pik} \sqrt{\normShort{\Curve'(u)}^2} \nonumber \\
                                           &= \frac{1}{2 \normShort{ \Curve'(u) }} \pderiv{}{\Pik} \normShort{\Curve'(u)}^2 \nonumber \\
                                           &= \frac{1}{2 \normShort{ \Curve'(u) }} \sum_{j=1}^{3N} \pderiv{}{\Pik} \left( C'_j(u) \right) ^2 \nonumber \\
                                           &= \frac{1}{2 \normShort{ \Curve'(u) }} \pderiv{}{\Pik} \left( C'_k(u) \right) ^2 \nonumber \\
                                           &= \frac{1}{2 \normShort{ \Curve'(u) }} 2 C'_k(u) \pderiv{C'_k(u)}{\Pik}
\end{align}
where again, the sum collapses because $\Pik$ only affects the $k$-th dimension of the B-spline curve derivative.
$\inlinepderiv{C'_k(u)}{\Pik}$ can be calculated with standard algorithms.\cite{heuer2018a,piegl1997a}


\section*{References}


\providecommand{\refin}[1]{\\ \textbf{Referenced in:} #1}







\end{document}